\documentclass[a4paper]{jpconf}  
\usepackage[dvips]{graphicx}
\usepackage{dcolumn}   
\usepackage{bm}        
\usepackage{amssymb}   
\usepackage[utf8]{inputenc}
\usepackage{mathrsfs}
\usepackage{amsmath}
\usepackage[breaklinks=true,colorlinks=true]{hyperref}
\hypersetup{colorlinks=true,citecolor=blue,linkcolor=blue,urlcolor=blue}
\usepackage{textcomp}
\usepackage{subfigure}
\usepackage{color}

\begin{document}

\title{Vibrations of thick domain walls: How to avoid no-go theorem}

\author{Vakhid A. Gani}

\address{Department of Mathematics, National Research Nuclear University MEPhI\\ (Moscow Engineering Physics Institute), Moscow 115409, Russia}

\address{Theory Department, Institute for Theoretical and Experimental Physics\\ of National Research Centre ``Kurchatov Institute'', Moscow 117218, Russia}

\ead{vagani@mephi.ru}

\begin{abstract}
The asymptotic properties of the stability potentials of kinks with power-law tails are discussed. In particular, in cosmology such kinks can describe ``thick'' domain walls. The discrete part of the domain wall excitation spectrum, the profile of which is described by a kink solution with one or both power-law asymptotics, cannot contain levels other than zero (translational) mode. Nevertheless, it can be shown that scenarios are quite possible when long-lived localized vibrations will be excited on the domain wall. This, in turn, can affect the processes of interaction of two or more domain walls.
\end{abstract}

\section{Introduction and motivation}

Kink solutions with power-law asymptotic behavior are of growing interest. Such solutions arise in $(1+1)$-dimensional field-theoretic models with specific potentials. In particular, kinks with power-law asymptotics are present in models with polynomial potentials having minima of the fourth order or higher \cite{Christov.PRD.2019,Blinov.arXiv.2020}.

Many interesting and important results have been obtained. In particular, estimates for the interaction forces of kinks are found \cite{Christov.PRL.2019,Manton.JPA.2019,dOrnellas.JPC.2020,Campos.arXiv.2020}, improved initial conditions for the study of kink-antikink interactions are developed \cite{Christov.PRD.2019}. In addition, the study of collisions of kinks having power-law tails has started \cite{Christov.arXiv.2020,Belendryasova.CNSNS.2019}. In particular, in such collisions, resonance phenomena --- escape windows --- were observed. The appearance of escape windows is known to be a consequence of the resonant energy exchange between the translational modes of kinks (their kinetic energy) and their vibrational modes \cite{Belova.UFN.1997}. However, as it was discovered earlier, there are no vibrational modes in the excitation spectrum of any kink with a power-law asymptotic behavior. It would seem that something is wrong here. However, as will be shown below, there are possible solutions to this problem.

\section{Power-law asymptotics of kink}

Consider a model with polynomial or non-polynomial potential, which at $\varphi\approx\varphi_0^{}$ can be written as
\begin{equation}\label{eq:potential_approx}
    V(\varphi) \approx \frac{1}{2}\left(\varphi-\varphi_0^{}\right)^{2k} v(\varphi_0^{}),
\end{equation}
where $v(\varphi_0^{})>0$ is a constant, and $k$ is a positive integer. Below we study kink solution $\varphi_{\rm K}^{}(x)$ which interpolates between two neighboring vacua of the potential \eqref{eq:potential_approx}, and $\lim\limits_{x \to +\infty} \varphi_{\rm K}^{}(x)=\varphi_0^{}$. It can be shown \cite{Blinov.arXiv.2020} that at $k>1$ the kink $\varphi_{\rm K}^{}(x)$ has power-law right asymptotics:
\begin{equation}\label{eq:kink_asymptotics}
    \varphi_{\rm K}^{}(x) \approx \varphi_0^{} - \frac{A_k^{}}{x^{1/(k-1)}} \quad \mbox{at} \quad x\to+\infty,
\end{equation}
where
\begin{equation}\label{eq:kink_asymptotics_coefficient}
     A_k^{} = \left[\left(k-1\right)\sqrt{v(\varphi_0^{})}\right]^{1/(1-k)}.
\end{equation}
Now, before discussing the properties of the excitation spectrum of kink with power-law asymptotics, we should make a brief digression on the kink's stability potential.

\section{Kink's stability potential}

The standard procedure of obtaining the kink's stability potential includes the following steps (see, e.g., \cite[Sec.~2]{Bazeia.EPJC.2018} for more details):

\begin{itemize}

\item add a small perturbation $\delta\varphi(x,t)$ to the static kink $\varphi_{\rm K}^{}(x)$, i.e.\ we write
\begin{equation}
\varphi(x,t) = \varphi_{\rm K}^{}(x) + \delta\varphi(x,t);
\end{equation}
\item assume that
\begin{equation}
\delta\varphi(x,t) = \eta(x)\cos\:\omega t;
\end{equation}
\item substituting into the equation of motion, we obtain the eigenvalue problem
\begin{equation}
\hat{H}\eta(x) = \omega^2\eta(x),
\end{equation}
where
\begin{equation}
\hat{H} = -\frac{d^2}{dx^2} + U(x).
\end{equation}

\item The function $U(x)$ reads
\begin{equation}\label{eq:stability_potential}
U(x) = \left.\frac{d^2 V}{d\varphi^2}\right|_{\varphi_{\rm K}^{}(x)}
\end{equation}
and it is called {\it kink's stability potential}.

\end{itemize}

(Note that in many cases the kink $\varphi_{\rm K}^{}(x)$ is known only in the implicit form $x=x_{\rm K}^{}(\varphi)$.)

\section{Properties of the stability potential}

The following properties of the stability potential of a kink with one or both power-law asymptotics can be proven (detailed discussion of these properties will be given in a separate publication).

\bigskip

\begin{enumerate}

\item The stability potential is volcano-like, if both (left and right) kink's asymptotics are power-law; it is symmetric or not, depending on the symmetry of the kink.

\item Asymptotic behavior of the kink's stability potential is universal:
\begin{equation}
    U(x) \approx \frac{B_k^{}}{x^2} \quad \mbox{at} \quad x\to+\infty,
\end{equation}
where
\begin{equation}
    B_k^{} = \frac{k\left(2k-1\right)}{\left(k-1\right)^2 v(\varphi_0^{})}.
\end{equation}
(Here we assume that the kink has asymptotics \eqref{eq:kink_asymptotics}.)

\item No-go theorem for vibrational modes:

there is only the zero mode in the discrete part of the kink's excitation spectrum, which lies on the boundary of the continuous spectrum.

\end{enumerate}

\section{Ways to avoid the above no-go theorem}

\bigskip

\subsection{The first option}

It would seem that the absence of vibrational modes in the excitation spectrum of a kink with power-law asymptotics excludes the possibility of resonance phenomena in the kink-antikink scattering. However, a closer examination of the problem reveals a loophole associated with the features of the kink's stability potential. If the kink has exponential asymptotics on the left and power-law asymptotics on the right, then the corresponding stability potential is an asymmetric potential well with edges at different heights. In this case, $\lim\limits_{x\to-\infty}U(x)>\lim\limits_{x\to+\infty}U(x)=0$. This means that if we consider `kink+antikink' configuration, we can expect that the potential determining the excitation spectrum of such a system will have the form of a potential well of variable width, depending on the distance between the kink and the antikink. The discrete part of the spectrum in such a well can have several eigenvalues. The situation is similar to that observed in the $\varphi^6$ model \cite{Dorey.PRL.2011}. In connection with kinks with power-law asymptotics, the possibility of such a scenario was also indicated in \cite{Belendryasova.CNSNS.2019}. Recall that, despite the absence of a vibrational mode in the excitation spectrum of a kink, resonance phenomena have already been observed in collisions of kinks with one exponential and one power-law asymptotics \cite{Christov.arXiv.2020,Belendryasova.CNSNS.2019}. In these cases, the colliding kinks were facing each other exactly with power-law tails.

\subsection{The second option}

\begin{figure}[h!]
\begin{center}
  \includegraphics[width=0.49\textwidth]{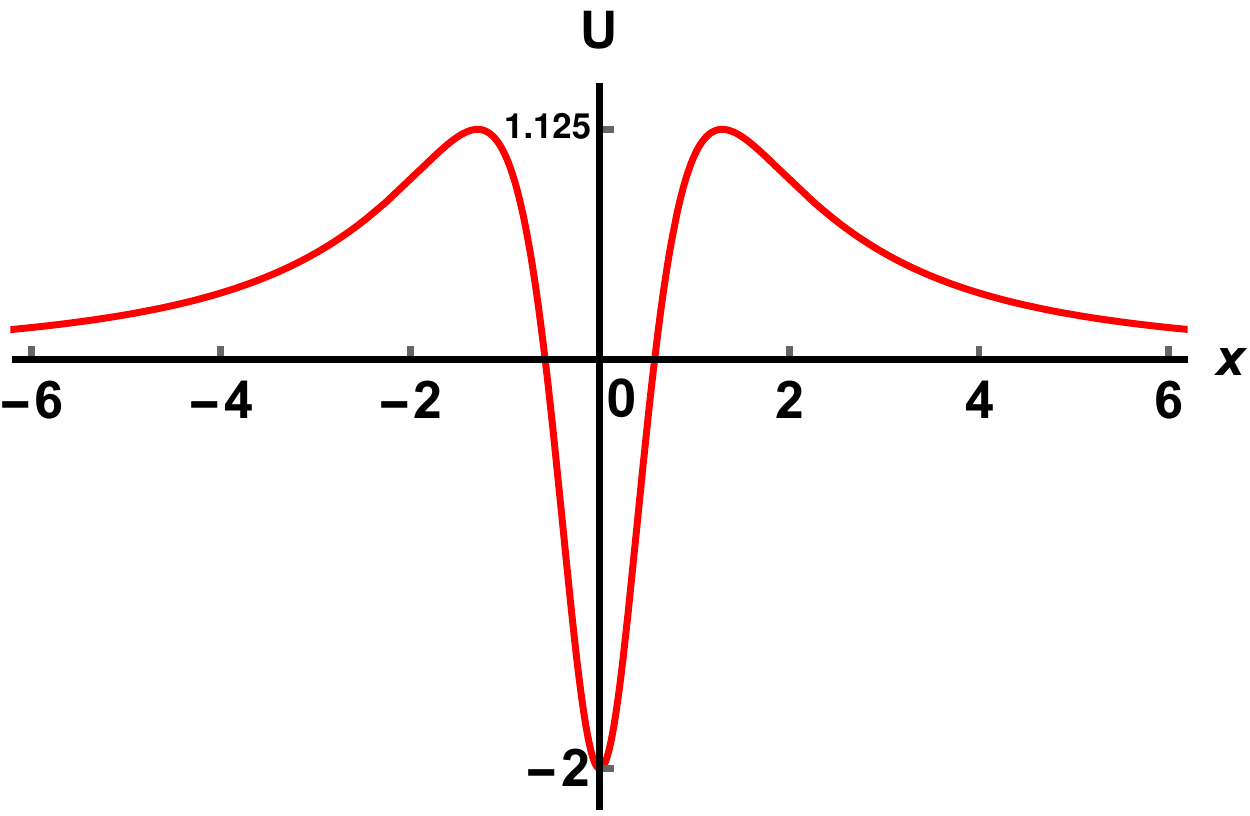}\label{fig:QMPkink}
  \caption{Stability potential \eqref{eq:stability_potential} for a symmetric kink with power-law asymptotics: the potential is $V(\varphi) = \displaystyle\frac{1}{2}\cos^4\varphi$, the corresponding kink is $\varphi_{\rm K}^{}(x) = \text{arctg}\: x$. }
  \label{fig:fig1}
\end{center}
\end{figure}

The stability potential of a kink with both power-law asymptotics is volcano-like (moreover, in the case of a symmetric kink, this potential will also be symmetric), see example in Fig.~\ref{fig:fig1}.
Such form of the potential means that above the boundary of the continuous spectrum (this boundary corresponds to the zero eigenvalue, which is the translational mode of the kink) there can exist ``quasi-discrete levels''. Frequencies of those ``quasi-discrete levels'' are formally in the continuous spectrum but could correspond to long-lived excitations localized on the kink. However, the question of the possibility of the existence of such levels is still open.

\section{Concluding remarks}

The study of the excitation spectra of kinks with power-law asymptotics is an important problem. As already mentioned, in collisions of kinks and antikinks, which have power-law asymptotics, resonance phenomena, i.e., escape windows, were observed. However, a theory describing the resonant exchange between kinetic energy and vibrational modes is still lacking. It is hoped that the study of the asymptotic behavior, as well as other features of the stability potentials of kinks with power-law tails, will make it possible to create a complete theory of the scattering of such kinks.

The results described above are preliminary. It is expected that the work will be continued and these and some other results will be published soon.

\section*{Acknowledgements}

The research was supported by the Russian Foundation for Basic Research for their support under Grant No.\ 19-02-00930.

The author also acknowledges the support of the MEPhI Academic Excellence Project.

\end{document}